\documentclass[epj,nopacs]{svjour}
\usepackage{graphics}

\begin{document}
\title{Novel constraints on light elementary particles and
extra-dimensional physics from the Casimir effect}
\author{R.~S.~Decca\inst{1} \and D.~L\'{o}pez\inst{2} \and
E.~Fischbach\inst{3} \and
G.~L.~Klimchitskaya\inst{4,}\thanks{On leave from North-West Technical University,
{\protect \\} Millionnaya St. 5, St.Petersburg,
191065, Russia}  \and
D.~E.~Krause\inst{5,3} \and
V.~M.~Mostepanenko\inst{4,}\thanks{On leave from Noncommercial
Partnership ``Scientific {\protect \\} Instruments'', Tverskaya St. 11, Moscow,
103905, Russia}
}                     
\institute{Department of Physics, Indiana University-Purdue
University Indianapolis, Indianapolis, Indiana 46202, USA  \and
Bell Laboratories, Lucent Technologies, Murray Hill,
New Jersey 07974, USA \and
Department of Physics, Purdue University, West Lafayette, Indiana
47907, USA \and
Center of Theoretical Studies and Institute for Theoretical Physics,
Leipzig University,
Augustusplatz 10/11, 04109, {\protect \\}Leipzig, Germany \and
Physics Department, Wabash College, Crawfordsville, Indiana 47933,
USA
}
\date{Received: date / Revised version: date}
%
\abstract{
We present supplementary information on the recent indirect
measurement
of the Casimir pressure between two parallel plates using a
micromachined oscillator.
The equivalent pressure between the plates is obtained by means of the
proximity force approximation after measuring the force gradient between
a gold coated sphere and a gold coated plate.
The data are compared with a new theoretical
approach to the thermal Casimir force based on the use of the Lifshitz
formula, combined with a generalized plasma-like dielectric permittivity
which takes into account interband transitions of core electrons.
The theoretical Casimir pressures calculated using the new approach are
compared with those computed in the framework of the previously used
impedance approach and also with the Drude model approach. The latter is shown to be
excluded by the data at a 99.9\% confidence level within a wide separation
range from 210 to 620\,nm.
The level of agreement between the data and theoretical approaches based
on the generalized plasma model, or the Leontovich surface impedance, is used to
set stronger constraints on the Yukawa forces predicted from the exchange
of light elementary particles and/or extra-dimensional physics. The resulting
constraints are the strongest in the interaction region from 20 to 86\,nm
with a largest improvement by a factor of 4.4 at 26\,nm.
%
} 
\authorrunning{R.~S.~Decca et al.}
\titlerunning{Novel constraints on extra-dimensional physics from the
Casimir effect}
\maketitle
\section{Introduction}
\label{intro}
It is well known that there is little or no experimental
confirmation for many predictions of unified field theories,
supersymmetry, supergravity, or string theory. Direct
experimental tests for many of these predictions require
accelerators of very high energies which will be not available in
the foreseeable future. For this reason any
non-accelerator tests of the predictions of new physics
beyond the standard model attract the serious attention of
both experimentalists and theorists.

One of the most intriguing predictions made by many extensions
of the standard model is the existence of light and massless
elementary particles which arise as a result of some spontaneously
(or weakly dynamically) broken symmetry. Beams of such particles
can penetrate through thick matter with practically no interaction.
This makes it difficult to investigate these particles
using the usual laboratory setups
of elementary particle physics. There is, however, an alternative
way to investigate light elementary particles and their
interactions by using table-top laboratory experiments.
These experiments utilize the fact that
the exchange of such particles between atoms belonging to two
different macrobodies can generate a new long-range force in
addition to the commonly known electromagnetic and gravitational
interactions. For example, the exchange of predicted light bosons,
such as scalar axions, graviphotons, hyperphotons, dilatons and
moduli among others (see, e.g. \cite{1,2,3,4,5}) generates a
Yukawa potential. The simultaneous exchange of two photons, two
massless scalars or massless pseudoscalars, and the exchange of
a massless axion or a massless neutrino-antineutrino pair leads
to power-law interactions with different powers \cite{6,7,8,9,10,11}.
Coincidently, a Yukawa correction to Newtonian gravity is predicted
in extra-dimensional physics with compact extra dimensions and low
energy compactification scale of order of 1\,TeV \cite{12,13,14,15}.
Furthermore, some brane theories contain exactly the standard model at
low energy \cite{15a}.
For models of non-compact but warped extra dimensions, power-law
corrections to the Newtonian gravitational law have been predicted
\cite{16}.
The cosmological constant generated in such models may be of
the correct order
of magnitude as suggested by observations \cite{16aa}.
Direct experimental signatures of strings and branes
are discussed in \cite{16a}.

Experimental constraints on hypothetical long-range interactions arising
from both light elementary particles and large extra dimensions can be
obtained from precise force measurements between macrobodies. For electrically
neutral test bodies the dominant background force at separations greater than
$10^{-5}\,$m is the gravity. At shorter separations the dominant
forces are the van der Waals and Casimir forces caused by fluctuations of
the electromagnetic field \cite{16b}. During the past few years a number of new
experiments have been performed to measure small forces between macrobodies and
to obtain stronger constraints on hypothetical long-range interactions
(which are also referred to as the ``fifth force'' \cite{1}). Thus, in
sub-millimeter gravity experiments stronger constraints on Yu\-ka\-wa corrections
to the Newtonian gravitational force for ran\-ges $\sim 10^{-4}\,$m and $\sim 10^{-5}\,$m
have been obtained \cite{17,18,19,20,21,22}. In a series of experiments
measuring the Casimir force between gold coated test bodies the constraints
on Yukawa-type interactions in a sub-micrometer range have been strengthened up to
$10^{4}$ times \cite{23,24,25,26,27,28,29,30,31,32,33}.

This paper exploits the results of the most precise recent determination
of the Casimir pressure between two parallel gold coated plates
using a micromechanical torsional oscillator. This is the third in a series
of experiments using a micromechanical oscillator for precise Casimir force
measurements. Results of the first two experiments were published in
\cite{32,33} (previously a similar technique was used to demonstrate the actuation
of a micromechanical device by the Casimir force \cite{34}). A brief discussion
of the results of the third experiment, and a description of the main improvements,
as compared with the previous two experiments, is contained in \cite{35}. Here we present
additional experimental details related to the experiment \cite{35} which were
not discussed in the first publication, including the resistivity
measurements and tests of the linearity of the oscillator used. The focus of this
paper is a comparison of the experimental data with a recently proposed new
theoretical approach to the thermal Casimir force \cite{36} which is applicable
to all experiments regardless of the separation between the interacting bodies.
Within this framework, we first present a precise fit
of the tabulated optical data \cite{37} for the imaginary part of the dielectric
permittivity of gold within a wide frequency region.
The fit is obtained using a set of six oscillators representing interband
transitions in gold.
We then compare this fit with
a previously known fit based on DESY data \cite{38,39}. Our theoretical
approach based on the
Lifshitz formula is found to be in very good agreement with the measured
results. The same measured results are also compared with an alternative approach
to the theory of the thermal Casimir force \cite{40}, which approach is found to be
excluded by our measurements at a confidence level of 99.9\%. The level of
agreement between our theory and experimental data is used to set constraints
on Yukawa-type corrections to Newtonian gravity originating from the exchange of
light hypothetical elementary particles and/or extra-dimensional physics.
The resulting constraints are several times stronger than those derived from
previous experiments. We also reanalyze constraints following
\cite{31} from experiment \cite{26} (in \cite{31} the confidence level of our
 results was not determined). As a consequence, the interaction region where
the constraints from the present experiment are the strongest is
widened. Special attention is paid to minor deviations between experiment and
theory at the shortest separations. Although these deviations are inside the
error bars and thus not statistically meaningful, we present an analysis of
various explanations for them.

The plan of this paper is as follows: in Sec.~2 we present a brief description of
the experimental setup and measurement results with an emphasis on novel aspects
not described previously in \cite{32,33,35}. Sec.~3 is devoted to the comparison of
experimental data with different theoretical approaches including our new
approach in \cite{36}. The new precise oscillator fit of the optical data for gold
is also presented here. Sec.~4 contains constraints on hypothetical Yukawa
interactions following from the level of agreement of data with
theory, and includes a comparison with
constraints obtained from earlier experiments. In Sec.~5 we present
our conclusions and discussion.

\section{Experimental setup and measurement results}
\label{sec:1}

One component of our setup is an Au-coated sapphire sphere attached to an optical
fiber. The thickness of the Au coating on
the sphere is $\Delta_g^{\!(s)}=180\,$nm, and the
radius of the coated sphere is $R=151.3\pm 0.2\,\mu$m. The sphere is placed
at a separation $z$ above a micromachined oscillator consisting of a heavily
doped, Au-coated polysilicon plate (the thickness of the coating is
$\Delta_g^{\!(p)}=210\,$nm)
suspended at two opposite points by serpentine springs. This plate can rotate
under the influence of the Casimir force $F(z)$ acting between the sphere and
the plate. The rotation angle is measured by the change of the capacitance
between the plate and two independently contacted polysilicon electrodes
located under it (details of the setup are described in \cite{32,33}). The
micromachined oscillator and the sphere with a fiber were mounted inside a can
with magnetic damping vibration isolation, where a pressure below
$10^{-4}\,$torr was maintained.

In this experiment a dynamic measurement mode was employed. For this purpose
the vertical separation between the sphere and the plate was varied harmonically,
$\tilde{z}(t)=z+A_z\cos(\omega_rt)$, where $\omega_r$ is the resonant angular
frequency of the oscillator in the presence of the sphere. The magnitude of
the amplitude $A_z\approx 2\,$nm was chosen in such a way that the oscillator
exhibited a linear response.
In the presence of the Casimir force $F(z)$
the resonant frequency $\omega_r$ is shifted
relative to the natural angular frequency of the oscillator
$\omega_0=2\pi\times(713.25\pm 0.02)\,$Hz determined in the absence of the
sphere. In the linear regime this shift can be found using
\cite{32,33,34,35}
\begin{equation}
\omega_r^2=\omega_0^2\left[1-\frac{b^2}{I\omega_0^2}
\frac{\partial F(z)}{\partial z}\right],
\label{eq1}
\end{equation}
\noindent
where $b$ is the lever arm between the axis of plate rotation and the projection
on the plate of the closest point of the sphere, $I$ is the moment of inertia
of the oscillator, and $b^2/I=(1.2432\pm 0.0005)\,\mu\mbox{g}^{-1}$.

The actual measured quantity in this experiment is the change of the
resonant frequency of the oscillator, $\omega_r-\omega_0$, under
the influence of the Casimir force $F(z)$ acting between the sphere and the
plate. Using (\ref{eq1}), the experimental data for $\omega_r-\omega_0$
obtained at different separation distances can be transformed into
$\partial F(z)/\partial z$. It is less useful, however, to recover the
force $F(z)$ between a sphere and a plate using the force gradient.
A better avenue is given by using the proximity force approximation (PFA)
\cite{41,42,43}
\begin{equation}
F(z)=2\pi RE(z),
\label{eq2}
\end{equation}
\noindent
where $E(z)$ is the Casimir energy per unit area between two infinitely
large parallel plates composed of the same materials as the sphere and the
plate. Differentiating with respect to $z$ and taking into account that
the Casimir pressure between the two parallel plates is
\begin{equation}
P(z)=-\frac{\partial E(z)}{\partial z},
\label{eq3}
\end{equation}
\noindent
one arrives to the expression
\begin{equation}
P(z)=-\frac{1}{2\pi R}\frac{\partial F(z)}{\partial z}.
\label{eq4}
\end{equation}
\noindent
From equations (\ref{eq1}) and (\ref{eq4}) one can immediately convert the experimental
data into data for the Casimir pressure between
two parallel plates.
This is in fact the so-called indirect measurement \cite{51}
of the pressure.
Note that in \cite{44}, where the
configuration of two parallel plates was actually used in the experimental setup,
the directly measured quantity was also the frequency shift due to the
Casimir pressure proportional to $\partial P(z)/\partial z$.
The pressure $P(z)$ was then recovered using the data for its derivative.

The calibration of absolute separations between the plate and the sphere was
performed by the application of voltages in a manner analogous  to that
reported in \cite{32,33}. The use of a two-color fiber interferometer
\cite{45} and a $\approx 7$\% improvement in vibration noise yielded an
error of only 0.2\,nm in a distance $z_{meas}$ between the end of the fiber
and the stationary reference. As a result, for every repetition of the
Casimir pressure measurement we were able to reposition our sample to
within $\Delta z_{meas}=0.2\,$nm. Finally the absolute separations $z$
between the sphere and the plate were measured with an absolute error
$\Delta z=0.6\,$nm determined at 95\% confidence \cite{33}.

The indirect measurements of the Casimir pressure $P_j(z_i)$ were repeated at
practically the same separations $z_i$ ($1\leq i\leq 293$)  33 times
($1\leq j \leq 33$). The mean values of the experimental Casimir pressure
\begin{equation}
\bar{P}(z_i)=\frac{1}{33}\sum\limits_{j=1}^{33}
P_j(z_i)
\label{eq5}
\end{equation}
\noindent
are plotted in Fig.~1 as a function of separation over the entire measurement
range from $z_1=162.03\,$nm to $z_{293}=745.98\,$nm.
As an example, a few mean Casimir pressures $\bar{P}(z_i)$ at different separations are
presented in column ($a$) of Table~1.
In this measurement the random experimental error is much smaller than the systematic error.
Specifically, using Student's $t$-distribution
\cite{49a} with a number of degrees of
freedom $f=32$, and choosing $\beta=0.95$ confidence, we obtain
$p=(1+\beta)/2=0.975$, and $t_p(f)=2.0$. This leads to the random
experimental error
\begin{equation}
\Delta^{\! \rm rand}P^{\rm exp}(z_i)=s(z_i)t_p(f),
\label{eq6}
\end{equation}
\noindent
where $s(z)$ is the variance of the mean for the pressure
\begin{equation}
s^2(z_i)=\frac{1}{1056}\sum\limits_{j=1}^{33}
\left[P_j(z_i)-\bar{P}(z_i)\right]^2.
\label{eq7}
\end{equation}
\noindent
The random error in (\ref{eq6}) reaches a maximum value
equal to 0.46\,mPa at $z=162\,$nm, decreases to 0.11\,mPa at $z=300\,$nm,
and maintains this value up to $z=746\,$nm.

\begin{figure}[b]
\vspace*{-9cm}
\resizebox{0.75\textwidth}{!}{%
 \includegraphics{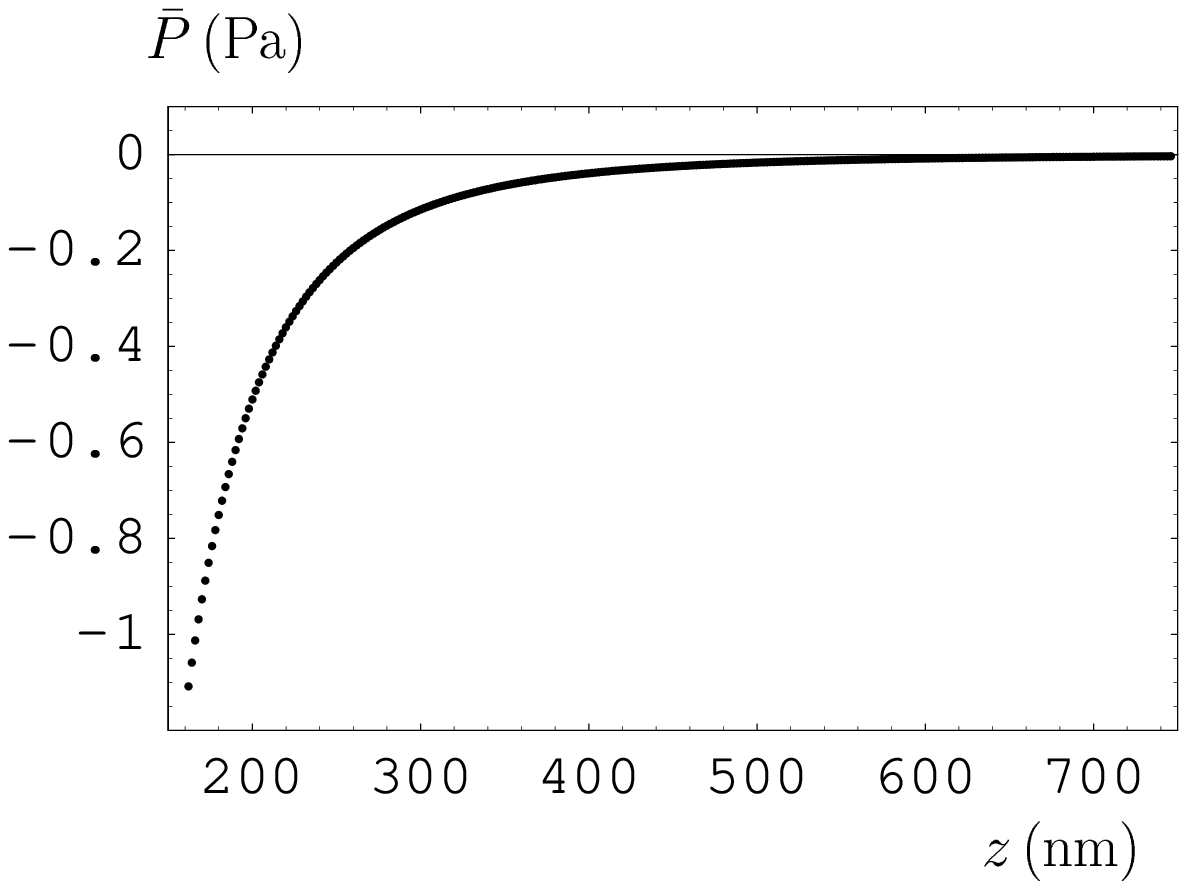}
}
\vspace*{-4.2cm}
\caption{Values of the mean Casimir pressure between two Au-coated plates
as a function of separation.}
\label{fig:1}       
\end{figure}
The systematic error of the pressure measurements in this experiment is
determined by the errors in the measurements of the resonance frequency,
radius of the sphere (these errors were indicated above), and also by
the error in using the PFA. Until 2006 the latter was not known with
certainty but estimated to be of order $z/R$ on the basis of dimensional
considerations \cite{43}. Recently, however, quantitative results on the
accuracy of PFA were obtained theoretically for the configuration of a
cylinder above a plate \cite{46,47} (the electromagnetic Casimir effect),
and for a sphere above a plate \cite{48,49} (the scalar Casimir effect). In addition
the validity of the PFA  was established
experimentally \cite{50} for a sphere above a plate.
In all cases at small separations the error in
using the PFA was shown to be less than $z/R$. However, in
our conservative error analysis we estimate this error with a safety
margin as $z/R$. By combining all the above systematic errors at 95\%
confidence using the statistical rules described in \cite{33}, we
obtain a systematic error equal to 2.12\,mPa at $z=162\,$nm. The systematic
error decreases to 0.44\,mPa at $z=300\,$nm, and then to 0.31\,mPa at
$z=746\,$nm. Finally we combine the resulting random and systematic
errors at a 95\% confidence  to arrive at the total
experimental error, $\Delta^{\! \rm tot}P^{\rm exp}(z)$,
approximately equal to the systematic error at all
separations considered.
Detailed information on the statistical methods used in our error
analysis can be found in \cite{33,51}.
As a result, the total relative experimental
error $\Delta P^{\rm exp}(z)/|\bar{P}(z)|$ varies from 0.19\% at
$z=162\,$nm, to 0.9\% at $z=400\,$nm, and to 9.0\% at $z=746\,$nm.
Hence this is the most precise experiment on the Casimir effect performed
up to date.
\begin{table}
\caption{Magnitudes of the mean experimental Casimir pressures
$\bar{P}$ (column $a$) at different separations $z$ compared with the
magnitudes of the theoretical pressures $P^{\rm th}$ computed using the
generalized plasma model approach (column $b$), the Leontovich surface
impedance approach (column $c$), the Drude model approach (column $d$),
and with the half-width, $\Xi$, of the 95\% confidence interval for
$P^{\rm th}-\bar{P}$ (column $e$). All pressures are given in mPa.}
\label{tab:1}       
\begin{center}
\begin{tabular}{llllll}
\hline\noalign{\smallskip}
$z\,$(nm) & {\ \ }$a$ & {\ \ }$b$ & {\ \ }$c$ & {\ \ }$d$ & {\ }$e$  \\
\noalign{\smallskip}\hline\noalign{\smallskip}
162 & 1108.4 & 1098.4 & 1094.2 & 1076.2 & 21.2 \\
166 & 1012.7 & 1007.1 & 1002.7 & 985.40 & 19.0 \\
170 & 926.85 & 923.71 & 919.56 & 902.96 & 17.1 \\
180 & 751.19 & 750.58 & 747.06 & 732.14 & 13.3 \\
190 & 616.00 & 616.71 & 613.70 & 600.28 & 10.5 \\
200 & 510.50 & 511.26 & 508.70 & 496.62 & 8.40 \\
250 & 225.16 & 225.71 & 224.45 & 217.11 & 3.30 \\
300 & 114.82 & 114.87 & 114.18 & 109.48 & 1.63 \\
350 & 64.634 & 64.574 & 64.176 & 61.004 & 0.98 \\
400 & 39.198 & 39.096 & 38.850 & 36.617 & 0.69 \\
450 & 25.155 & 25.034 & 24.874 & 23.247 & 0.54 \\
500 & 16.822 & 16.785 & 16.678 & 15.456 & 0.47 \\
550 & 11.678 & 11.669 & 11.595 & 10.654 & 0.42 \\
600 & 8.410 & 8.365 & 8.312 & 7.573 & 0.39 \\
650 & 6.216 & 6.151 & 6.113 & 5.522 & 0.38 \\
700 & 4.730 & 4.626 & 4.598 & 4.118 & 0.36 \\
746 & 3.614 & 3.620 & 5.598 & 3.198 & 0.35 \\
\noalign{\smallskip}\hline
\end{tabular}
\end{center}
\end{table}

Several additional measurements and tests were performed in order to compare
the experimental data with theory in a conclusive manner.
In order to include the effects of surface roughness in theoretical
computations of the Casimir pressure we have investigated the topography of
the metallic coatings both on the plate ($p$) and on the sphere ($s$) using
an AFM probe in tapping mode.
All AFM scans were squares with sizes ranging from $0.5\times
0.5\,\mu$m to $10\times 10\,\mu$m. The information obtained was
indistinguishable. In the case of a sphere the surface curvature
was taken into account. For this purpose the image was
planarized, and then the roughness analysis performed. For a
typical scan of $5\times 5\,\mu$m the effect of curvature is about
40\,nm.
From AFM images of the surfaces, the fraction
of each surface area $v_i^{(p,s)}$ with height $h_i^{(p,s)}$ was determined.
It was found that for the sphere ($1\leq i\leq K^{(s)}=106$) $h_i^{(s)}$
varies from 0 to 10.94\,nm and for the plate
($1\leq i\leq K^{(p)}=85$) $h_i^{(p)}$
varies from 0 to 18.35\,nm.
Here, the highest peaks on the sphere and on the plate are almost
of the same height as in the previous experiment of \cite{33}
(11.06\,nm and 20.65\,nm on the sphere and plate, respectively
\cite{33}). However, they are much lower than the highest peaks in
the experiment \cite{32}.
The respective zero roughness levels on the
sphere and on the plate, $H_0^{(s)}$ and $H_0^{(p)}$, are found from
\begin{equation}
\sum\limits_{i=1}^{K^{(p,s)}}\left[H_0^{(p,s)}-h_i^{(p,s)}\right]
v_i^{(p,s)}=0.
\label{eq8}
\end{equation}
\noindent
From (\ref{eq8}) using the roughness data one obtains $H_0^{(s)}=5.01\,$nm
and $H_0^{(p)}=9.66\,$nm. Note that precise measurements of
absolute separations $z$ discussed
above result in separations just between the zero roughness levels
determined in (\ref{eq8}).

Special tests were performed to investigate possible nonlinear behavior of the
oscillator under the influence of the Casimir force. First, the resonance
frequency $\omega_r$ observed under the excitation leading to a harmonically
varying separation with amplitude $A_z$ was compared with the resonance
frequency with no excitation (i.e., with separation varied just through the
thermal noise). When the amplitude $A_z$ was less than 4\,nm, no deviation
was observed between the measured resonance frequency and the thermal
resonance  frequency within the $\approx\,5\,$mHz noise. This was performed
at different separations (recall that the amplitude actually used in the experiment
was $A_z\approx 2\,$nm). The value of $A_z^{cr}$ at which
deviations are observed is a function of separation. For example, at
$z=199.8\,$nm $A_z^{cr}=4.5\,$nm, at $z=247.3\,$nm $A_z^{cr}=10.0\,$nm, and
at $z=302.4\,$nm,  $A_z^{cr}=15.0\,$nm. In all cases, when observed,
nonlinearities decrease the resonance frequency.

\begin{figure}[b]
\vspace*{-9cm}
\resizebox{0.75\textwidth}{!}{%
 \includegraphics{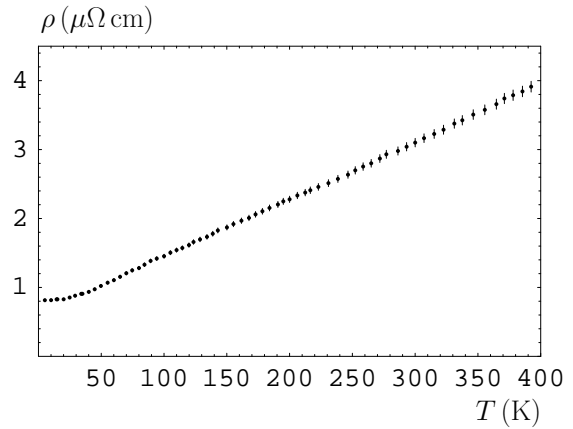}
}
\vspace*{-3.2cm}
\caption{Resistivity of the Au films (measured with an error of about 2\%) as a function
of temperature.}
\label{fig:2}       
\end{figure}
Another test performed was a check for the strength of the signal at different
harmonics of the excitation. The experiment was done with the excitation
$\tilde{z}(t)$ at the resonant frequency $\omega_r$. The checks were performed
with the excitations at frequencies $2\omega_r$ and $\omega_r/2$, but no
change in the response at $\omega_r$ was observed for $A_r<4\,$nm.
For larger amplitudes the results were consistent with what was observed
in the first test. These tests all verify that the oscillator
was in fact operating in a linear regime for our measurements.

It is significant that the comparison of experimental data with theory of the
Casimir force requires knowledge of the optical and electronic parameters
of the Au layers. In previous experiments all of these parameters, including the
plasma frequency $\omega_p$ and relaxation parameter $\gamma(T)$, were taken
from tables \cite{37}. For a more conclusive comparison of this experiment with
different theoretical approaches, we measured the resistivity $\rho$ of the
Au films as a function of temperature in the region from
$T_1=3\,$K to 400\,K. These measurements were performed
 using a four probe approach on Au films of the
same thickness which were  deposited at the same time as the Au deposition on the
oscillator, and on the same substrates.
The samples were approximately 1\,mm long and $10\,\mu$m wide. The resistivity
of each sample was found by taking into account its geometrical factor
with an error of about 2\% arising from the errors in measuring of the
sample's geometry. The experimental data for the resistivity versus temperature
are presented in Fig.~2. These data at $T\gg T_D/4$ (where $T_D=165\,$K is the
Debye temperature for Au) were fitted to a straight line \cite{52}
\begin{equation}
\rho(T)=\frac{4\pi}{\omega_p^2\tau(T)}=\frac{4\pi v_F}{\omega_p^2l(T)}
=\frac{CT}{\omega_p^{3/2}}.
\label{eq9}
\end{equation}
\noindent
Here $\tau(T)=l(T)/v_F$ is the relaxation time,  $l(T)\sim T$ is the mean free
path of an electron, $v_F\sim \omega_p^{1/2}$ is the Fermi velocity,
and $C=\mbox{const}$. The fit results in
$C/\omega_p^{3/2}=(8.14\pm 0.16)\,\mbox{n}\Omega\,\mbox{cm}\,\mbox{K}^{-1}$.
On the other hand, using the resistivity data for pure Au as a function of
temperature \cite{53} and the previously used
value of the plasma frequency $\tilde{\omega}_p=9.0\,$eV \cite{37,54}
we obtain $C/\tilde{\omega}_p^{3/2}=8.00$. As a result we find for
the Au film used in our experiment
$\omega_p=(8.9\pm0.1)\,$eV.
Here, the absolute error of 0.1\,eV arises from the errors of the
resistivity measurements.
Some of the theoretical approaches to the
thermal Casimir force require a knowledge of the relaxation parameter.
The smooth Drude extrapolation of the imaginary part of the Au dielectric
permittivity, given by the tabulated optical data \cite{37}, yields
the relaxation parameter at room temperature
$\gamma=0.0357\,$eV (which compares with $\tilde{\gamma}=0.035\,$eV
used in previous work \cite{32,33,54}).

\section{Comparison of experimental data with different theoretical
approaches to the thermal Casimir force}
\label{sec:2}

The theoretical description of both the van der Waals and Casimir pressures
between planar plates at temperature $T$ in thermal equilibrium
is given by the Lifshitz  formula \cite{55a}
\begin{eqnarray}
&&P(z)=-\frac{k_BT}{\pi}
\sum\limits_{l=0}^{\infty}\left(1-\frac{1}{2}\delta_{l0}\right)
\int_{0}^{\infty}k_{\bot}dk_{\bot}q_l
\nonumber \\
&&
\phantom{aa}\times
\left\{\left[r_{\rm TM}^{-2}(\xi_l,k_{\bot})e^{2q_lz}-1\right]^{-1}
\right.
\label{eq10} \\
&&
\phantom{aaaa}+\left.
\left[r_{\rm TE}^{-2}(\xi_l,k_{\bot})e^{2q_lz}-1\right]^{-1}\right\}.
\nonumber
\end{eqnarray}
\noindent
Here $k_{\bot}=|\mbox{\boldmath$k$}_{\bot}|$ is the magnitude of the wave vector
component in the plane of the plates, $q_l^2=k_{\bot}^2+\xi_l^2/c^2$,
$\xi_l=2\pi k_BTl/\hbar$ are the Matsubara frequencies,
$l=0,\,1,\,2,\,\ldots\,$,
$\delta_{lm}$ is Kronecker's delta symbol,
and $k_B$ is the Boltzmann constant.
The reflection coefficients for two independent polarizations of the
electromagnetic field (the transverse magnetic and transverse electric)
are defined as
\begin{equation}
r_{\rm TM}(\xi_l,k_{\bot})=\frac{\varepsilon_lq_l-k_l}{\varepsilon_lq_l+k_l},
\quad
r_{\rm TE}(\xi_l,k_{\bot})=\frac{k_l-q_l}{k_l+q_l},
\label{eq11}
\end{equation}
\noindent
where
\begin{equation}
k_l=\sqrt{\varepsilon_l\frac{\xi_l^2}{c^2}+k_{\bot}^2},
\quad
\varepsilon_l=\varepsilon(i\xi_l),
\label{eq12}
\end{equation}
\noindent
and $\varepsilon(\omega)$ is the frequency-dependent dielectric permittivity
of the plates.

Note that (\ref{eq10}) is the expression for a plate of infinite thickness.
Using the Lifshitz formula for layered structures \cite{43}, it is easy to see
that for Au layer thicknesses larger than 150\,nm (as in our case) at, e.g.,
$z=400\,$nm the error due to the replacement of a layer with a semispace is
less than 0.003\%.

It is known that there is some controversy concerning the
application of (\ref{eq10}), (\ref{eq11}) to real metals. These controversies arise
from different approaches to the calculation of the zero-frequency ($l=0$)
term in (\ref{eq10}). For real materials (Au for instance)
$\varepsilon(i\xi_l)$ is usually found through the Kramers-Kronig
relation
\begin{equation}
\varepsilon(i\xi_l)=1+\frac{1}{\pi}\mbox{P}\int_{-\infty}^{\infty}
\frac{\omega\varepsilon^{\prime\prime}(\omega)}{\omega^2+\xi_l^2}\,d\omega,
\label{eq13}
\end{equation}
\noindent
where $\varepsilon^{\prime\prime}(\omega)$ is the imaginary part of the
dielectric permittivity and the integral is taken as a principal value.
Optical data for $\varepsilon^{\prime\prime}(\omega)$ are available
within some restricted frequency region \cite{37}, and it is common to
smoothly extrapolate available data to lower frequencies using the
imaginary part of the Drude model dielectric permittivity
\begin{equation}
\varepsilon^{\prime\prime}(\omega)=
\frac{\omega_p^2\gamma}{\omega(\omega^2+\gamma^2)}.
\label{eq14}
\end{equation}
\noindent
If such an extrapolation is performed down to lower frequencies, including zero
frequency \cite{40}, the use of the resulting $\varepsilon(i\xi_l)$ in the
Lifshitz theory leads to a violation of the Nernst heat theorem for
perfect crystal lattices \cite{55}, and the Casimir pressures calculated using
(\ref{eq10}) are in contradiction with experiment \cite{32,33,35}. Because of
this, two other approaches to the determination of  $\varepsilon(i\xi_l)$
were proposed in the literature. According to the plasma model approach \cite{56,57},
the tabulated optical data are not used and $\varepsilon(i\xi_l)$ is found
from the free electron plasma model
\begin{equation}
\varepsilon(i\xi_l)=1+\frac{\omega_p^2}{\xi_l^2}.
\label{eq15}
\end{equation}
\noindent
According to the impedance approach \cite{58,59}, the reflection coefficients
(\ref{eq11}) are expressed in terms of the Leontovich surface impedance
$Z(\omega)$ instead of the dielectric permittivity. The contributions of all
Matsubara frequencies with $l\geq 1$ are obtained from tabulated
optical data extrapolated by the Drude model using a relation
$Z(i\xi_l)=1/\sqrt{\varepsilon(i\xi_l)}$. This leads to approximately the same
calculation results as the use of the dielectric permittivity. As to the
contribution of zero Matsubara frequency, it is obtained using the impedance
of infrared optics, and is different from that obtained using the Drude
model (a discussion of different approaches can be found in \cite{60,61}).
Note that the impedance approach was used for the first comparison of
the measurement data of this experiment with theory \cite{35}.

Although the plasma model and impedance approaches are in agreement with
thermodynamics, neither can be considered as universally valid. The plasma
model approach completely neglects dissipation. Because of this, it is
in agreement with measured data only for experiments \cite{32,33,35}
performed at separations larger than the plasma wavelength $\lambda_p$.
As for the impedance approach, it is not applicable to short-separation
experiments \cite{26} because when $z<\lambda_p$ the Leontovich impedance
boundary conditions become invalid due to the violation of the inequality
$|Z(\omega)|\ll 1$.

Recently \cite{36} a new approach to the thermal Casimir force between
real metals was proposed which is equally applicable at both small and large
separations. This approach is based on the use of the generalized
plasma-like dielectric permittivity
\begin{equation}
\varepsilon(\omega)=1-\frac{\omega_p^2}{\omega^2}+
\sum\limits _{j=1}^{K}
\frac{f_j}{\omega_j^2-\omega^2-ig_j\omega}\, ,
\label{eq16}
\end{equation}
\noindent
which takes into account the interband transitions of core electrons.
Here $\omega_j\neq 0$ are the resonant frequencies of the core electrons,
$g_j$ are the respective relaxation frequencies, $f_j$ are the oscillator
strengths, and $K$ is the number of oscillators. Note that the term
$-\omega_p^2/\omega^2$ on the right-hand side of (\ref{eq16}) describes free
electrons and leads to a purely imaginary current. This contribution to
$\varepsilon(\omega)$ is entirely real and does not include dissipation.
Importantly, the oscillator term on the right-hand side of (\ref{eq16}) does
not include the oscillator with zero resonant frequency $\omega_0=0$,
which is equivalent to the Drude dielectric function, i.e., it does not
describe conduction electrons but only core electrons. This term
incorporates dissipation due to interband transitions.

In \cite{36} the Lifshitz theory together with the dielectric permittivity
(\ref{eq16}) was used to calculate the thermal Casimir force in a
short-separation experiment \cite{26}, and the experimental results were found to be
in good agreement with theory. For this purpose the oscillator parameters of
Au in (\ref{eq16}) were taken from \cite{38,39} where they were found using the
3-oscillator model fitted to old DESY data. Below we compare the 3-oscillator fit
of \cite{38,39} with the more complete data set of \cite{37} and perform a more exact
6-oscillator fit. The resulting oscillator parameters are used to calculate the
Casimir pressure in the most precise experiment described in the previous
section.

The Kramers-Kronig relation (\ref{eq13}) was derived \cite{62} for dielectric
permittivities which were regular or which had a first order pole at zero frequency.
For the dielectric permittivity (\ref{eq16}) which has a second order pole at
$\omega=0$, the Kramers-Kronig relation expressing $\varepsilon(i\omega)$
in terms of $\varepsilon^{\prime\prime}(\omega)$ is the following \cite{36}:
\begin{equation}
\varepsilon(i\xi_l)=1+\frac{1}{\pi}\mbox{P}\int_{-\infty}^{\infty}
\frac{\omega\varepsilon^{\prime\prime}(\omega)}{\omega^2+\xi_l^2}\,d\omega
+\frac{\omega_p^2}{\xi_l^2}.
\label{eq17}
\end{equation}

In the tables of \cite{37} the most complete data are collected for real,
$n_1(\omega)$, and imaginary, $n_2(\omega)$, parts of the complex refraction
index of Au in the frequency region from 0.125\,eV to 9919\,eV
($1\,\mbox{eV}=1.519\times 10^{15}\,$rad/s).
From these data, the imaginary part of the Au dielectric permittivity is expressed
as $2n_1(\omega)n_2(\omega)$. To obtain the contribution of core electrons
to the dielectric permittivity, we consider the difference
\begin{equation}
\varepsilon_{Au}^{\prime\prime}(\omega)=2n_1(\omega)n_2(\omega)-
\frac{\tilde{\omega}_p^2\tilde{\gamma}}{\omega(\omega^2+\tilde{\gamma}^2)},
\label{eq18}
\end{equation}
\noindent
where in accordance with (\ref{eq14}) the subtracted term  approximately
describes the contribution of free conduction electrons to optical data.
\begin{figure}[b]
\vspace*{-9cm}
\resizebox{0.75\textwidth}{!}{%
 \includegraphics{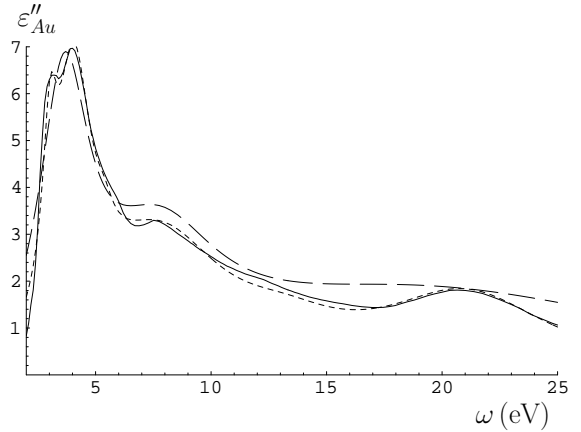}
}
\vspace*{-3.2cm}
\caption{Tabulated optical data for the imaginary part of Au dielectric permittivity
\cite{37} (with the contribution of conduction electrons subtracted) are shown by the
solid line. The oscillator fits are shown as long-dashed line \cite{38,39}
(DESY data, 3 oscillators) and as short-dashed line (6 oscillators).}
\label{fig:3}       
\end{figure}
In Fig.~\ref{fig:3} the quantity $\varepsilon_{Au}^{\prime\prime}$ is plotted
as a function of $\omega$ within the frequency region from 2.0\,eV to 25\,eV
(solid line). For $\omega<2\,$eV the dielectric permittivity
is determined by free conduction electrons, and for $\omega>2.5\,$eV
there is already practically no contribution from conduction electrons, and
$\varepsilon_{Au}^{\prime\prime}(\omega)\approx 2n_1(\omega)n_2(\omega)$.
The upper limit of the region under consideration is determined by the
frequencies contributing to the Casimir pressure (\ref{eq10}). Even at the
shortest separation considered, $z=160\,$nm, the characteristic frequency
$\Omega_c=c/(2a)\approx 0.62\,$eV. Bearing in mind that even for precise
computations of the pressure it is sufficient to take into account the
contribution from Matsubara frequencies up to $15\Omega_c$, setting the upper
limit of our region equal to 25\,eV is more than adequate.

The solid line in Fig.~\ref{fig:3} was fitted to the imaginary part of dielectric
permittivity (\ref{eq16})
\begin{equation}
\varepsilon^{\prime\prime}(\omega)=\sum\limits_{j=1}^{K}
\frac{f_jg_j\omega}{(\omega_j^2-\omega^2)^2+g_j^2\omega^2}
\label{eq19}
\end{equation}
\noindent
with $K=6$ oscillators. The resulting set of oscillator parameters $f_j$, $\omega_j$ and
$g_j$ is presented in Table~2. In Fig.~\ref{fig:3} the imaginary part of permittivity
calculated using the analytic expression (\ref{eq19}) is shown by the short-dashed
line. In the same figure the 3-oscillator fit from \cite{38,39} is shown as the
long-dashed line. As is seen in Fig.~\ref{fig:3}, the short-dashed line based on
the 6-oscillator fit better reproduces the actual data than does the long-dashed line
using the 3-oscillator fit.
\begin{table}[h]
\caption{The oscillator parameters for Au in equations (\ref{eq16}) and
(\ref{eq19}) found here from the 6-oscillator fit to the tabulated optical
data in \cite{37}.}
\label{tab:2}       
\begin{center}
\begin{tabular}{cccc}
\hline\noalign{\smallskip}
$j$ & $\omega_j\,$(eV) & $g_j\,$(eV) & $f_j\,(\mbox{eV}^2)$  \\
\noalign{\smallskip}\hline\noalign{\smallskip}
1 & 3.05 & 0.75 & 7.091 \\
2 & 4.15 & 1.85 & 41.46 \\
3 & 5.4{\ } & 1.0{\ } & 2.700 \\
4 & 8.5{\ } & 7.0{\ } & 154.7 \\
5 & 13.5 & 6.0{\ } & 44.55 \\
6 & 21.5 & 9.0{\ } & 309.6 \\
\noalign{\smallskip}\hline
\end{tabular}
\end{center}
\end{table}

The Casimir pressure $P_L(z)$ at all separations of interest was computed
using the Lifshitz theory in equations
(\ref{eq10})--(\ref{eq12}) and (\ref{eq16}) with $\omega_p=8.9\,$eV, as
determined for our films in Sec.~2, and the oscillator parameters from Table~2.
For comparison with the experimental data, the values of $P_L(z)$ were geometrically
averaged over all possible separations between the rough surfaces
weighted with the fractions of the total area occupied by each separation,
as discussed in Sec.~2. This results in the theoretical Casimir pressures
taking surface roughness into account via the equation
\begin{eqnarray}
&&
P^{\rm th}(z_i)=\sum\limits_{k=1}^{K^{(s)}}\sum\limits_{j=1}^{K^{(p)}}
v_k^{(s)}v_j^{(p)}
\nonumber \\
&&\phantom{aaaa}\times
P_L\left(z_i+H_0^{(s)}+H_0^{(p)}-h_k^{(s)}-h_j^{(p)}\right).
\label{eq20}
\end{eqnarray}
\noindent
The pressures $P^{\rm th}(z_i)$ were computed at each experimental point $z_i$.
Note that (\ref{eq20}) takes into account the
combined (nonmultiplicative) effect of
nonzero temperature and finite conductivity on the one hand [this is incorporated
in $P_L(z)$ computed using the Lifshitz formula], and of surface roughness on the
other. The contributions of diffraction-type and correlation effects in the
roughness correction \cite{63,64}, which are not taken into account in
the geometrical averaging (\ref{eq20}), were shown to be negligible \cite{33}.
In this experiment the contribution of the roughness correction to the Casimir pressure
computed using (\ref{eq20}) is very small. For example, at $z=162\,$nm the roughness
correction contributes only 0.52\% of the total pressure. At separations
$z=170,\> 200$ and 350\,nm roughness contributes only 0.48, 0.35 and 0.13\%
of the Casimir pressure, respectively. The magnitudes of the computed theoretical
Casimir pressures at some experimental separations are listed in column ($b$)
of Table~1.

We now discuss the accuracy of our computations. One of the sources of the
theoretical errors is the sample-to-sample variation of the optical data for the
complex index of refraction. As was shown in \cite{33} (see also \cite{26}),
in our experiments the variation of the optical data leads to an uncertainty in the
magnitude of the Casimir pressure which is substantially smaller than 0.5\%.
To be conservative, we admit an uncertainty
as large as 0.5\% in the computations due to the use
of tabulated optical data over the entire measurement range.
There are claims in the literature \cite{65} that the theoretical computations of
the Casimir pressure between gold surfaces are burdened by up to 5\% errors due to
the use of different Drude parameters measured and calculated for different
samples. This is, however, irrelevant to our experiment. The hypothesis
that the magnitude of $\omega_p$ is much smaller than the value we have used above (i.e.,
$\omega_p=6.85\,$eV or 7.50\,eV, as suggested in \cite{65}) is rejected at high
confidence by our experiment, and by all previously performed measurements
of the Casimir force between Au surfaces.

The other possible source of theoretical errors is connected with the fact that we
compute the Casimir pressure at experimental separations which are
determined with an error
$\Delta z=0.6\,$nm \cite{66}. Noting that the dominant
theoretical dependence of the Casimir pressure is $z^{-4}$,
 one finds that the relative  error in the pressure is equal to
$4\Delta z/z$. This varies from 1.5\% at $z=160\,$nm to 0.32\% at $z=750\,$nm.
The other theoretical errors, e.g., arising from neglect of patch potentials or
spatial nonlocality, were analyzed in detail in \cite{33} and found to be  negligible.
By combining the above two theoretical errors discussed here and in the
previous paragraph at a 95\% confidence level using the
statistical procedure applicable to systematic errors described by a uniform
distribution \cite{33,51}, we obtain the total theoretical error
$\Delta^{\!\rm tot} P^{\rm th}(z)$
as a function of separation.
This error assumes a maximum value of 18.7\,mPa at $z=162\,$nm,
which is almost 9 times larger than the total experimental error. Note that in
a similar analysis in \cite{33}, one additional theoretical error due to the use
of PFA was included. In \cite{33} it was first combined with
the theoretical error due to sample-to-sample variation of the optical data,
with the result that
the distribution law of the combined quantity was not uniform. In this work,
however, the error due to the accuracy of PFA is included with the experimental
errors. Because of this, the total theoretical error is determined by only two
contributions. With the increase of separation to $z=300$, 400 and 746\,nm, the
total theoretical error decreases to 1.15, 0.34 and 0.024\,mPa, respectively.
The relative theoretical error $\Delta P^{\rm th}(z)/|P^{\rm th}(z)|$ assumes a
maximum value of 1.7\% at $z=162\,$nm. When the separation increases
to $z=300$, 400 and 746\,nm, the relative theoretical error decreases to 1.0,
0.86 and 0.65\%, respectively. This is mainly explained by the decreased role
of uncertainty in determining the separations.

We can now compare experiment and theory by considering the differences
$P^{\rm th}(z_i)-\bar{P}(z_i)$ at each experimental separation $z_i$. The confidence
interval for the quantity $P^{\rm th}(z_i)-\bar{P}(z_i)$ determined at 95\%
confidence probability is given by $\left[-\Xi_{0.95}(z_i),\Xi_{0.95}(z_i)\right]$
where the half-width of this interval can be found using the composition rule
\cite{33,51}
\begin{eqnarray}
&&
\Xi_{0.95}(z_i)=\min\left\{\Delta^{\! \rm tot}P^{\rm th}(z_i)+\Delta^{\!\rm tot}P^{\rm exp}(z_i),
\right.
\nonumber \\
&&
\phantom{aa}\left.
k_{0.95}^{(2)}\sqrt{\left[\Delta^{\! \rm tot}P^{\rm th}(z_i)\right]^2+
\left[\Delta^{\!\rm tot}P^{\rm exp}(z_i)\right]^2}\right\}.
\label{eq21}
\end{eqnarray}
\noindent
Here for two composed quantities $k_{0.95}^{(2)}=1.1$. The values
of the half-width of the confidence interval are listed in the last column of Table~1.

In Fig.~4a the differences $P^{\rm th}(z_i)-\bar{P}(z_i)$ at all experimental points
are shown as dots.
\begin{figure}[h]
\vspace*{0cm}
\resizebox{0.75\textwidth}{!}{%
 \includegraphics{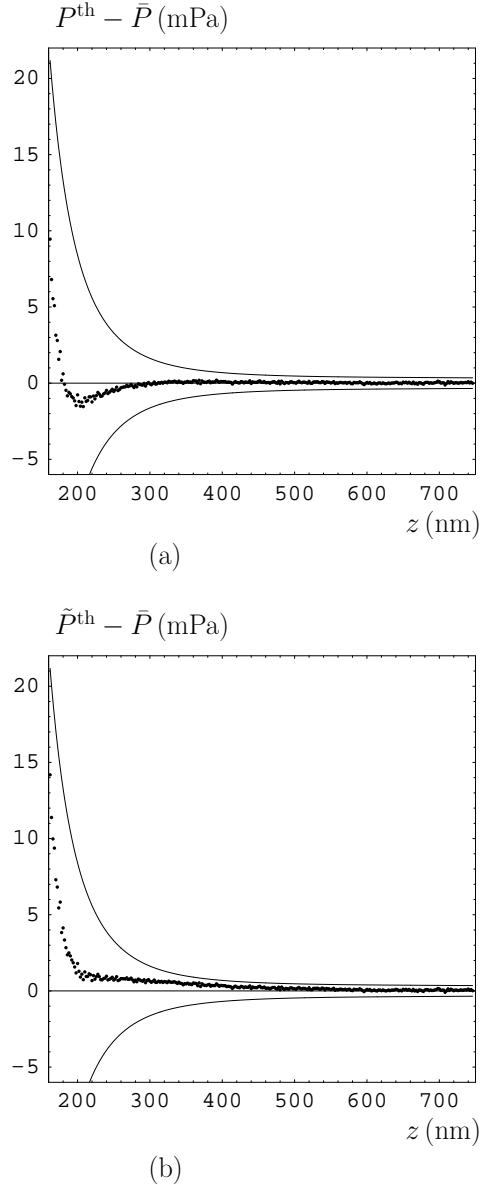}
}
\vspace*{-1cm}
\caption{Differences between theoretical Casimir pressures computed using the
generalized plasma model approach (a) and the Leontovich surface impedance approach
(b) and mean experimental Casimir pressures (dots) versus separation. Solid lines indicate
the limits of the 95\% confidence intervals.}
\label{fig:4}
\end{figure}
In the same figure the confidence interval $\left[-\Xi_{0.95}(z_i),\Xi_{0.95}(z_i)\right]$
at each $z$ is situated between the solid lines. As seen in the figure, all dots (and not
only 95\% of them as required by the rules of mathematical statistics) are well inside the
confidence interval at all separations considered. This means that the experimental data
are consistent with theory based on the
generalized plasma-like dielectric permittivity (\ref{eq16}),
and that in our conservative error analysis the errors (especially at short separations)
are overestimated. For comparison purposes in Fig.~4b we plot as dots the differences
$\tilde{P}^{\rm th}(z_i)-\bar{P}(z_i)$ where the experimental data are the same as
in Fig.~4a, but with $\tilde{P}^{\rm th}(z_i)$  computed as in \cite{35} using the Leontovich
surface impedance approach, with the Drude parameters $\omega_p=8.9\,$eV,
$\gamma=0.0357\,$eV. In Table~1, column ($c$) we present the magnitudes of the Casimir
pressures $\tilde{P}^{\rm th}(z_i)$ computed using the surface impedance approach at
different separations. As is seen in Fig.~4b, the impedance theoretical approach
is also consistent with the data. However, while in Fig.~4a there are practically no deviations
between experiment and theory at $z>350\,$nm, in Fig.~4b the deviations are noticeable
up to $z=450\,$nm. By comparing columns ($b$) and ($c$) in Table 1, we can conclude that the
differences between the two theoretical approaches do not exceed the magnitude of the
theoretical error.
The comparison of columns ($b$) and ($c$) with column ($a$) shows that at all separations
the approach using the generalized plasma-like model is in somewhat better agreement
with data than the surface impedance approach.
As is seen in Fig.~4a,b, the largest deviations between both theoretical
approaches and experimental data are at short separations from 162 to 200\,nm.
Although these deviations are not statistically meaningful, because they are well inside
the confidence interval, in Sec.~5 we will discuss possible reasons leading to the
deviations between experiment and theory at shortest separations.

\begin{figure}[b]
\vspace*{-8.5cm}
\resizebox{0.75\textwidth}{!}{%
 \includegraphics{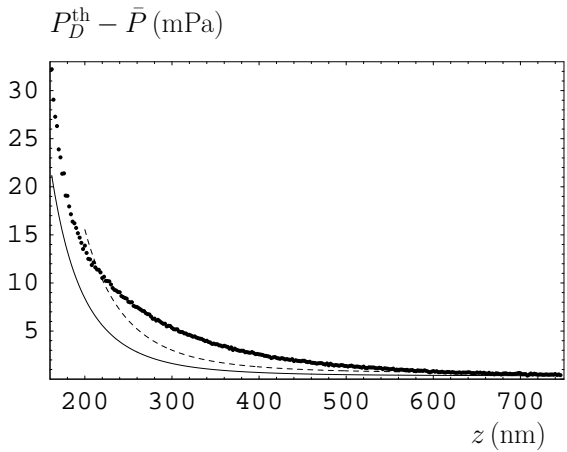}
}
\vspace*{-3.2cm}
\caption{Differences between theoretical Casimir pressures computed using the
Drude model approach and mean experimental Casimir pressures (dots) versus separation.
The solid line indicates the limits of the 95\% confidence intervals while the
dashed line indicates the limits of the 99.9\% confidence intervals.
}
\label{fig:5}       
\end{figure}
A completely different situation occurs when we compare the experimental data with
the alternative approach to the thermal Casimir force \cite{40} using the Drude model
to compute the contribution of the zero-frequency term in the Lifshitz formula.
To perform the comparison, we calculate the theoretical Casimir pressures in the
framework of \cite{40} with the refined values of the Drude parameters
$\omega_p=8.9\,$eV, $\gamma=0.0357\,$eV (all details of this approach and of computations
can be found in \cite{33}). The magnitudes of the resulting Casimir pressures
$P_{D}^{\rm th}$ at a few different separations are listed in Table~1, column ($d$).
In Fig.~5 we plot the differences $P_{D}^{\rm th}(z_i)-\bar{P}(z_i)$ at all
experimental separations. The confidence interval
$\left[-\Xi_{0.95}(z_i),\Xi_{0.95}(z_i)\right]$ at each $z_i$ is the same for all
theoretical approaches. Once again, the limits of the confidence
interval are denoted by the solid lines (in Fig.~5 only
one solid line is shown because practically all dots are above it). As is seen in Fig.~5,
the Drude model theoretical approach is experimentally excluded at a 95\% confidence
level within the whole measurement range from 162 to 746\,nm. This conclusion is
confirmed by the calculation data in Table~1. Subtracting the magnitudes of the
theoretical Casimir pressures, $|P_{D}^{\rm th}|$, in column ($d$) from the experimental
results, $|\bar{P}|$, in column ($a$), we obtain at all separations larger results than
the half-width of the confidence interval, $\Xi_{0.95}(z)$, given in column ($e$).

The wide gaps between the solid line and dots in Fig.~5 suggest that the Drude
model approach is actually excluded experimentally at an even higher confidence than 95\%.
To make this argument quantitative, we calculate the half-width of a confidence
interval at 99.9\% confidence from
\begin{equation}
\frac{\Xi_{0.999}(z)}{\Xi_{0.95}(z)}=
\frac{t_{(1+0.999)/2}(32)}{t_{(1+0.95)/2}(32)}\approx 1.85,
\label{eq22}
\end{equation}
\noindent
where $t_p(f)$ is the Student coefficient used in Sec.~2. The limits of the 99.9\%
confidence intervals obtained in (\ref{eq22}) are shown in Fig.~5 by the dashed line. As is seen
in Fig.~5, the differences $P_{D}^{\rm th}-\bar{P}$ are found outside of the 99.9\% confidence
interval at separations from 210 to 620\,nm. This conclusively demonstrates that our experiment
is irreconcilable with the Drude model approach to the thermal Casimir force.
At the same time, the approaches based on the generalized plasma-like dielectric permittivity,
and on the Leontovich surface impedance, are consistent with experiment.
Note that in our experiment the Drude model approach is excluded at separations
below $1\,\mu$m. In the proposed experiments \cite{67a,67b,67c} it is planned to test the
predictions of different theoretical approaches to the thermal Casimir force at separations
of about several micrometers.

\section{Constraints on Yukawa-type hypothetical interactions and
light elementary particles}
\label{sec:3}

As was mentioned in the Introduction, at separations between macroscopic bodies of about $1\,\mu$m
and less, the Ca\-si\-mir force is the dominant background force. From the level of agreement
between the experimental data for the Casimir pressure and Lifshitz theory (with a
generalized plasma-like permittivity in Sec.~3), one can constrain any additional
force which may coexist with the Casimir force. As noted in the Introduction, many
extensions of the standard model predict a Yukawa correction to the Newtonian potential
energy between two point masses $m_1$ and $m_2$ at a separation $r$,
given by \cite{10,14,15}
\begin{equation}
V(r)=-\frac{Gm_1m_2}{r}\left(1+\alpha e^{-r/\lambda}\right).
\label{eq23}
\end{equation}
\noindent
Here $G$ is the Newtonian gravitational constant, $\alpha$ is a dimensionless constant
characterizing the strength of the Yukawa interaction, and $\lambda$ is its range.

The total force acting between two parallel plates due to the potential (\ref{eq23}) can be
obtained by integration of (\ref{eq23}) over the volumes of the plates, and subsequent
negative differentiation with respect to $z$. In experiments measuring
the Casimir force the contribution of the gravitational force is very small and can be
neglected \cite{31,32}. Thus, in what follows we consider only the contribution from
the Yukawa term in (\ref{eq23}).

To find the Yukawa pressure for our setup
we should take into account the detailed structure of our
test bodies. (As was shown in Sec.~3, for the calculation of the Casimir pressure it
is possible to replace the Au coating films with Au semispaces and we need not consider the
underlying substrate.)
In the present experiment a sapphire sphere of density $\rho_s=4.1\,\mbox{g/cm}^{3}$ was
first coated with a layer of Cr of density $\rho_c=7.14\,\mbox{g/cm}^{3}$ and
thickness $\Delta_c=10\,$nm, and then with an external layer of gold of thickness
$\Delta_g^{\!(s)}=180\,$nm and density $\rho_g=19.28\,\mbox{g/cm}^{3}$. The Si plate of
thickness $L=3.5\,\mu$m, and density
$\rho_{Si}=2.33\,\mbox{g/cm}^{3}$ was also first coated with a layer of Cr of
$\Delta_c=10\,$nm thickness, and then with a layer of gold of
$\Delta_g^{\!(p)}=210\,$nm thickness. Under the conditions $z,\,\lambda\ll R$,
satisfied in this experiment, the equivalent Yukawa pressure between the two parallel
plates with the same layer structure as the above sphere and a plate is given by \cite{28,43}
\begin{eqnarray}
&&
P^{\rm hyp}(z)=-2\pi G\alpha\lambda^2 e^{-z/\lambda}
\label{eq24} \\
&&\phantom{a}\times
\left[\rho_g-(\rho_g-\rho_c)e^{-\Delta_g^{\!(s)}/\lambda}-
(\rho_c-\rho_s)e^{-(\Delta_g^{\!(s)}+\Delta_c)/\lambda}\right]
\nonumber \\
&&\phantom{a}\times
\left[\rho_g-(\rho_g-\rho_c)e^{-\Delta_g^{\!(p)}/\lambda}-
(\rho_c-\rho_{Si})e^{-(\Delta_g^{\!(p)}+\Delta_c)/\lambda}\right].
\nonumber
\end{eqnarray}
\noindent
We have verified that the surface roughness, as reported in Sec.~2, cannot
significantly affect the
magnitude of a hypothetical pressure with an interaction range longer than 10\,nm,
and hence can be
neglected. Because of this, there is no need to perform geometrical averaging as in
(\ref{eq20}) when calculating the Yukawa interaction.

According to Sec.~3, theories of the thermal Casimir force using the generalized plasma-like
permittivity or the Leontovich surface impedance are consistent with experimental data.
As was noted in Sec.~3, in our conservative analysis the errors [and consequently the
width of the confidence interval $2\Xi(z)$] are overestimated. The reason
for this is that we have
included the error due to the uncertainty of experimental separations $\Delta z$ in the analysis
of the theoretical errors. As a result, the theoretical pressures acquired an extra error of
${\approx}4\Delta z/z$ which led to enormous widening of the confidence interval at short
separations (see Figs.~4 and 5). This approach was useful in selecting among
different theories of the thermal Casimir force, and permitted us to exclude the
one based on the use of the Drude model at practically 100\% confidence. However, as is
clearly seen in Fig.~4a,b, the actual width of the confidence interval is much less
than that between the solid lines (recall that the actual confidence interval determined at
95\% confidence should contain about 95\% of the data dots but not all of them). It is easily seen that
if the theoretical error $4\Delta z/z$ due to uncertainties in experimental separations is
disregarded, the resulting more narrow confidence interval
$\left[-\tilde{\Xi}(z),\tilde{\Xi}(z)\right]$ still contains all dots representing
$P^{\rm th}(z)-\bar{P}(z)$ within the separation region from 180 to 746\,nm.
At a separation $z=180\,$nm, the half-width $\tilde{\Xi}=4.80\,$mPa. At typical separations
$z=200$, 250, 300, 350, 400 and 450\,nm $\tilde{\Xi}$ is equal to 3.30, 1.52, 0.84, 0.57,
0.45 and 0.40\,mPa, respectively. Thus, for $180\,\mbox{nm}\leq z\leq 746\,$nm
the magnitude of the hypothetical pressure should satisfy the inequality
\begin{equation}
|P^{\rm hyp}(z)|\leq\tilde{\Xi}(z).
\label{eq25}
\end{equation}
\noindent
Bearing in mind that the half-width of the confidence interval $\tilde{\Xi}(z)$ was
defined at a 95\% confidence, the same confidence also applies to the constraints following
from the inequality (\ref{eq25}).

We have performed a numerical analysis of equations (\ref{eq24}) and (\ref{eq25}) at
different separations and determined the resulting region of $(\lambda,\alpha)$-plane
where the inequality (\ref{eq25}) is satisfied, so that the existence of Yukawa interaction
is consistent with the level of agreement achieved between data on the measurement of the Casimir
force and relevant theory. The strongest constraints within the interaction region
$10\,\mbox{nm}\leq\lambda\leq 56\,$nm are obtained from the comparison of measurements
with theory at a separation $z=180\,$nm. With the increase of $\lambda$, the strongest
constraints on $\alpha$ were obtained from the consideration of larger separations.
Thus constraints in the regions $56\,\mbox{nm}\leq\lambda\leq 71\,$nm,
$71\,\mbox{nm}\leq\lambda\leq 89\,$nm,  $89\,\mbox{nm}\leq\lambda\leq 140\,$nm,
$140\,\mbox{nm}\leq\lambda\leq 220\,$nm and $220\,\mbox{nm}\leq\lambda\leq 500\,$nm
were obtained from the agreement between Casimir pressure measurements and theory
at separations $z=200$, 250, 300, 350 and 400\,nm, respectively.

\begin{figure}[b]
\vspace*{-6cm}
\resizebox{0.75\textwidth}{!}{%
 \includegraphics{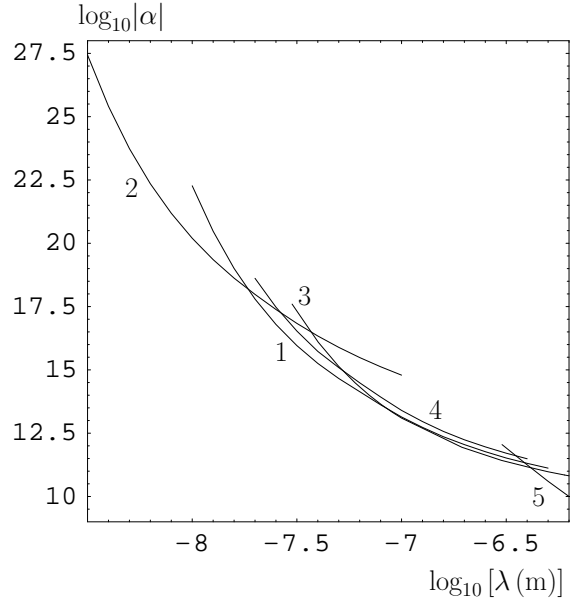}
}
\vspace*{-3cm}
\caption{Constraints on the strength of the Yukawa interaction versus interaction
range. Line 1 is obtained in this paper, line 2 was obtained in \cite{31} using
the Casimir force measurement of \cite{26} and adapted in this paper to the
accepted 95\% confidence level. Lines 3 and 4 were obtained in \cite{69}
and \cite{33}, respectively. Line 5 was obtained in the first reference of \cite{28}
using the Casimir force measurement \cite{23}. The region of the $(\lambda,\alpha)$
plane above each line is excluded and below the line is allowed.}
\label{fig:6}       
\end{figure}
The resulting constraints on $\alpha$ are plotted in Fig.~6 for different values of
$\lambda$ (line 1). The region in the $(\lambda,\alpha)$-plane above the line 1 is
excluded by the results of the Casimir pressure measurements compared with theory,
and below line 1 is allowed. For comparison, constraints from earlier experiments are
also shown in Fig.~6 in a similar manner.

Special attention should be paid to line 2 representing constraints following from
the short-separation experiment \cite{26} on the measurement of the Casimir force
between an Au-coated sphere and a plate using an atomic force microscope.
The constraints on a Yukawa hypothetical interaction following from that experiment
were obtained in \cite{31}, and later used  in \cite{32,33,35} for comparison with
constraints following from other experiments. However, in \cite{31} the level
of agreement between experiment and theory at zero temperature was described in terms
of the root-mean-square deviation which, as was recognized later in \cite{33}, is
not an appropriate quantity in strongly nonlinear situations. In addition, the
calculational scheme using the root-mean-square deviation does not permit us to
determine the confidence level of the results.

Here we reanalyze the experimental data of \cite{26} and compare them with theory using
the Lifshitz formula at laboratory temperatures in a sphere-plate configuration,
supplemented by the generalized plasma-like dielectric permittivity (\ref{eq16}).
The results of this reanalysis are expressed in terms of the confidence interval
$\left[-\Theta(z),\Theta(z)\right]$ determined at 95\% confidence for the differences
between theoretical and mean experimental Casimir forces, $F^{\rm th}(z)-\bar{F}(z)$.
This interval includes, in particular, the theoretical errors $3\Delta z/z$ arising due
to uncertainties of experimental separations in the sphere-plate configuration.
This interval cannot be narrowed as we did above in the case of the present experiment,
because the measurement in \cite{26} is
inherently noisier. For example, at $z=61.08\,$nm
the half-width of the confidence interval is $\Theta=31.6\,$pN, and with an increase of
separation up to 100.15 and 200.46\,nm it decreases to 9.17 and 7.20\,pN, respectively.
The resulting constraints at a 95\% confidence level are determined from
\begin{equation}
|F^{\rm hyp}(z)|\leq\Theta(z),
\label{eq26}
\end{equation}
\noindent
where $F^{\rm hyp}(z)$ is the Yukawa hypothetical force acting between an Au coated
sphere and a plate \cite{31,32}. These constraints are represented by line 2 in Fig.~6.
Note that the constraints given by line 2 are up to order of magnitude weaker than
those in \cite{31}, but they benefit from high confidence, and can be compared with
future work on the subject by using the same rigorous approach to the
comparison experiment with theory as proposed in \cite{33,67,68}.

The other lines in Fig.~6 are obtained from the Casimir-less experiment \cite{69}
(line 3), previous measurements of the Casimir pressure using the micromachined
oscillator \cite{33} (line 4), and in \cite{28} (the first paper) from a torsion pendulum
experiment \cite{23} (line 5). As is seen in Fig.~6, the resulting constraints
represented by line 1 are strongest within the interaction range
$20\,\mbox{nm}\leq\lambda\leq 86\,$nm with the largest improvement by a factor 4.4
at 26\,nm.
Note that further strengthening of the resulting constraints on $\alpha$ within
a submicrometer interaction range could provide important information concerning
predicted particles such as scalar axions, graviphotons, hyperphotons, dilatons,
and moduli among others. For such particles the interaction constant $\alpha$
could be much larger than unity. The same holds for theories based on extra-dimensional physics
with low-energy compactification scale where, for instance, for models with
three extra dimensions the predicted characteristic size of extra dimensions is
of about 5\,nm \cite{12,13}.

To conclude this section we briefly discuss possible reasons for the
observed deviations
between experiment and theory at the shortest separations shown in Fig.~4a,b.
These deviations are well inside the 95\% confidence interval determined for
$P^{\rm th}-\bar{P}$ and thus they are not statistically meaningful.
Nevertheless if we bear in mind that the deviations under consideration are several times
larger than the total experimental error, there may be some underlying physics
leading to the small discrepancies between experiment and theory. The most natural assumption
is that there is some undiscovered nonlinearity of the oscillator which results in an
additional systematic error at short separations. However, as discussed in Sec.~2,
special tests of the oscillator linearity have been performed which did not
reveal a nonlinear behavior for the amplitudes of sphere oscillations employed
in this experiment.
Another possible effect may be connected with some fine properties of interacting
surfaces determined, e.g., by correlation effects in surface roughness or by
patch potentials. However, as was analyzed in detail in \cite{33}, these effects are
negligibly small. Thus, even assuming enormously large patches due to monocrystals
with grain sizes ranging from $\sim\,300\,$nm (i.e., larger than the film thickness)
and to 25\,nm, the correction to the pressure due to patches at
$z=160\,$nm is only 0.42\,mPa (to be compared with the deviation between experiment and
theory of almost 10\,mPa, as in Fig.~4a).

We next consider the possibility that the deviation may be caused by
the Yukawa interaction (\ref{eq23}) with some appropriate
values of $\alpha$ and $\lambda$.
A simple calculation shows that the deviations between experiment and theory at short
separations would practically disappear if we allowed a Yukawa interaction with
$\alpha=5.0\times 10^{21}$ and $\lambda=10.4\,$nm. This interaction would
correspond to a point in Fig.~6 with log$\lambda=-7.98$ situated slightly below line 1.
However, such a point would lie above the point $\alpha=1.0\times 10^{20}$ with
the same $\lambda$ on line 2, which implies that
the assumed Yukawa interaction is excluded
by the AFM experiment \cite{26}. Bearing in mind that in the above we have
reanalyzed the results of \cite{26} at a 95\% confidence level using modern
methods of comparison between experiment and theory, this is strong evidence
against the existence of a single Yukawa interaction with
$\alpha=5.0\times 10^{21}$ and $\lambda=10.4\,$nm.
However, this analysis does not necessarily exclude the possible existence of more than one
Yukawa, or other interactions having a different spatial dependence.
A more decisive conclusion about the presence of hypothetical interactions can be
obtained through a repetition of the experiment described in Sec.~2 using a
Si plate but with no covering metallic layer. Noting that the density of
Au is in 8.3 times larger than the density of Si, and that the Casimir pressure between Au
and Si is approximately 1.5 times smaller than between Au and Au, the deviation
caused by the Yukawa interaction should practically disappear if the Au coated plate
is replaced with a Si plate.

\section{Conclusions and discussion}
\label{sec:4}

In this paper we have presented additional details on the recent
experimental determination of the Casimir pressure between two parallel plates
using a micromachined oscillator. This experiment incorporates several improvements
over all previous measurements. In particular, the measurements over a wide
separation range were repeated many times at practically the same points for
each repetition. This permits us to substantially reduce the random error, and to
make it much smaller than the systematic error for the first time in Casimir
force measurements. Also the plasma frequency of the Au films
was determined using the measured temperature dependence of their resistivity.

The resulting experimental data were compared with a new theoretical approach
to the thermal Casimir force using the Lifshitz theory
incorporating a generalized plasma-like
dielectric permittivity which takes into account the interband transitions.
For this purpose a new oscillator fit of the tabulated optical data for the
imaginary part of the dielectric permittivity of Au was performed which is more
exact than a previously used fit based on DESY data. The new theoretical approach
was also compared with the previously known approach using the Leontovich
surface impedance, and with the alternative approach using the Drude model.
The Drude model approach was excluded experimentally at a 99.9\% confidence level
over a wide separation range.

One of the main aims of this paper is the application of the Casimir effect to obtain
stronger constraints on hypothetical long-range interactions and light
elementary particles. We have reanalyzed the previously known constraints
from the measurement of the Casimir force between an Au-coated sphere and
a plate using modern methods of comparison of experiment and theory at
high confidence. We have also used the resulting level of agreement between
the measurements of the Casimir pressure and the new theory to strengthen constraints on
the hypothetical Yukawa-type interaction. The new constraints obtained above are the
strongest within the interaction range from 20 to 86\,nm, with the largest
improvement by a factor 4.4. These results are relevant for the verification of
different theoretical predictions made on the basis of unified field theories
beyond the standard model, and of extra-dimensional physics. We have also
discussed some possible reasons for small deviations between experiment and
theory at the shortest separations considered. It was shown that although these
systematic
deviations are not statistically significant, the fact that we have
no explanation for them at present suggests that
further experimental and theoretical work
is required to elucidate their nature.

\begin{acknowledgement}
{\it Acknowledgements.}
R.S.D. acknowledges NSF support through Grant No. CCF--0508239. E.F. was supported
in part by DOE under Grant No. DE--AC02--76ER071428. G.L.K. and V.M.M. are
grateful to Purdue University for kind hospitality and financial support.
They were also partially supported by DFG grant 436\.RUS\,113/789/0--3.
\end{acknowledgement}



\begin{thebibliography}{99}
\bibitem{1}
E.~Fischbach, C.~L.~Talmadge,
{\it The Search for Non-Newtonian Gravity}
(Springer-Verlag, New York, 1999).
\bibitem{2}
   E.~G.~Adelberger,  B.~R.~Heckel,
 A.~E.~Nelson,
{Ann. Rev. Nucl. Part. Sci.} {\bf 53}, 77 (2003).
\bibitem{3}
Y.~Fujii, Int. J. Mod. Phys. A {\bf 6}, 3505 (1991).
\bibitem{4}
   E.~G.~Adelberger,  B.~R.~Heckel, C.~W.~Stubbs,
 W.~F.~Rogers,
{Ann. Rev. Nucl. Part. Sci.} {\bf 41}, 269 (1991).
\bibitem{5}
S.~Dimopoulos, G.~F.~Giudice,
{Phys. Lett. B} {\bf 379}, 105 (1996).
\bibitem{6}
J.~Sucher, G.~Feinberg,
in {\it Long-Range Casimir Forces}, eds. F.~S.~Levin
and D.~A.~Micha (Plenum, New York, 1993).
\bibitem{7}
S.~D.~Drell, K.~Huang,
{ Phys. Rev.} {\bf 91}, 1527 (1953).
\bibitem{8}
F.~Ferrer, J.~A.~Grifols,
{ Phys. Rev. D} {\bf 58}, 096006 (1998).
\bibitem{9}
G.~Feinberg, J.~Sucher,
{ Phys. Rev.} {\bf 166}, 1638 (1968).
\bibitem{10}
E.~Fischbach,
Ann. Phys. (N.Y.) {\bf 247}, 213 (1996).
\bibitem{11}
 S. D. H. Hsu, P. Sikivie, Phys.
Rev. D {\bf 49}, 4951 (1994)
\bibitem{12}
I.~Antoniadis,
N.~Arkani-Hamed, S.~Dimopoulos,  G.~Dvali,
Phys. Lett. B {\bf 436}, 257 (1998).
\bibitem{13}
N.~Arkani-Hamed, S.~Dimopoulos,  G.~Dvali,
Phys. Lett. B {\bf 429}, 263 (1998);
Phys. Rev. D {\bf 59}, 086004 (1999).
\bibitem{14}
A.~Kehagias, K.~Sfetsos,
Phys. Lett. B {\bf 472}, 39 (2000).
\bibitem{15}
E.~G.~Floratos, G.~K.~Leontaris,
Phys. Lett. B {\bf 465}, 95 (1999).
\bibitem{15a}
C.~Kokorelis,
Nucl. Phys. B {\bf 677}, 115 (2004).
\bibitem{16}
L.~Randall, R.~Sundrum,
Phys. Rev. Lett. {\bf 83}, 3370 (1999);
{\bf 83}, 4690 (1999).
\bibitem{16aa}
A.~A.~Saharian,
Phys. Rev. D {\bf 74}, 124009 (2006).
\bibitem{16a}
I.~Antoniadis,
Int. J. Mod. Phys. A {\bf 21}, 1657 (2006).
\bibitem{16b}
M.~Kardar and R.~Golestanian,
Rev. Mod. Phys. {\bf 71}, 1233 (1999).
\bibitem{17}
G.~L.~Smith, C.~D.~Hoyle, J.~H.~Gundlach, E.~G.~Adelberger,
B.~R.~Heckel,  H.~E.~Swanson,
Phys. Rev. D {\bf 61}, 022001 (1999).
\bibitem{18}
C.~D.~Hoyle, U.~Schmidt, B.~R.~Heckel, E.~G.~Adelberger,
J.~H.~Gundlach, D.~J.~Kapner,  H.~E.~Swanson,
Phys. Rev. Lett. {\bf 86}, 1418 (2001).
\bibitem{19}
J.~C.~Long, H.~W.~Chan, A.~B.~Churnside, E.~A.~Gulbis,
M.~C.~M.~Varney,  J.~C.~Price,
Nature {\bf 421}, 922 (2003).
\bibitem{20}
J.~Chiaverini, S.~J.~Smullin, A.~A.~Geraci, D.~M.~Weld,
 A.~Kapitulnik,
Phys. Rev. Lett. {\bf 90}, 151101 (2003).
\bibitem{21}
S.~J.~Smullin, A.~A.~Geraci, D.~M.~Weld, J.~Chiaverini,
S.~Holmes, A.~Kapitulnik,
Phys. Rev. D {\bf 72}, 122001 (2005).
\bibitem{22}
D.~J.~Kapner, T.~S.~Cook, E.~G.~Adelberger, J.~H.~Gundlach,
B.~R.~Heckel, C.~D.~Hoyle,  H.~E.~Swanson,
Phys. Rev. Lett. {\bf 98}, 021101 (2007).
\bibitem{23}
S.~K.~Lamoreaux, { Phys. Rev. Lett.}
{\bf 78}, 5 (1997); {\bf 81}, 5475(E) (1998).
\bibitem {24}
U.~Mohideen, A.~Roy,
{ Phys. Rev. Lett.}
{\bf 81}, 4549 (1998);
G.~L.~Klimchitskaya, A.~Roy, U.~Mohideen,  V.~M. Mos\-te\-panenko,
{ Phys. Rev. A}
{\bf 60}, 3487 (1999).
\bibitem {25}
A.~Roy, C.-Y.~Lin,  U.~Mohideen,
{ Phys. Rev. D}
{\bf 60}, 111101(R) (1999).
\bibitem{26}
B.~W.~Harris, F.~Chen,  U.~Mohideen,
Phys. Rev. A {\bf 62}, 052109 (2000);
F.~Chen,
G.~L.~Klimchitskaya, U.~Mohideen,
 V.\ M.\ Mos\-te\-pa\-nen\-ko,
Phys. Rev. A
{\bf 69}, 022117 (2004).
\bibitem{27}
T.~Ederth, Phys. Rev. A {\bf 62}, 062104 (2000).
\bibitem{28}
M.~Bordag, B.~Geyer, G.~L.~Klimchitskaya,
V.\ M.\ Mos\-te\-pa\-nen\-ko,
{Phys. Rev. D}  {\bf 58}, 075003 (1998);
{\bf 60}, 055004 (1999);
{\bf 62}, 011701(R) (2000).
\bibitem{29}
J.~C.~Long, H.~W.~Chan,  J.~C.~Price,
Nucl. Phys. B {\bf 539}, 23 (1999).
\bibitem{30}
V.~M.~Mostepanenko, M.~Novello,
{Phys. Rev. D}
{\bf 63}, 115003 (2001).
\bibitem{31}
E.~Fischbach, D.~E.~Krause, V.~M.~Mostepanenko,  M.~Novello,
{Phys. Rev. D}
{\bf 64}, 075010 (2001).
\bibitem{32}
R.~S.~Decca, E.\ Fischbach, G.\ L.\ Klimchitskaya, D.\ E.\ Krause,
D.\ L\'opez,  V.\ M.\ Mostepanenko,
Phys. Rev. D {\bf 68}, 116003 (2003).
\bibitem{33}
R.~S.~Decca, D.~L\'opez, E.~Fischbach, G.\ L.\ Klimchitskaya,
D.\ E.\ Krause, V.\ M.\ Mostepanenko,
Ann Phys. (N.Y.) {\bf 318}, 37 (2005).
\bibitem{34}
H.~B.~Chan, V.~A.~Aksyuk, R.~N.~Kleiman, D.~J.~Bishop,  F.~Capasso,
Science {\bf 291}, 1941 (2001);
Phys. Rev. Lett. {\bf 87}, 211801 (2001).
\bibitem{35}
R.~S.~Decca, D.~L\'opez, E.~Fischbach, G.\ L.\ Klimchitskaya,
D.\ E.\ Krause,  V.\ M.\ Mostepanenko,
Phys. Rev. D {\bf 75}, 077101 (2007).
\bibitem{36}
G.~L.~Klimchitskaya, U.~Mohideen,  V.~M.~Mostepanenko,
J. Phys. A: Math. Theor. {\bf 40}, 339(F) (2007).
\bibitem{37}
{\it Handbook of Optical Constants of
Solids}, ed. E.\ D.\ Palik
(Academic, New York, 1985).
\bibitem{38}
V.~A.~Parsegian, G.~H.~Weiss,
J. Colloid Interface Sci. {\bf 81}, 285 (1981).
\bibitem{39}
V.~A.~Parsegian,
{\it Van der Waals Forces: A Handbook for Biologists,
Chemists, Engineers, and Physicists}
(Cambridge University Press, Cambridge, 2005).
\bibitem{40}
M.~Bostr\"{o}m, B.~E.~Sernelius,
Phys. Rev. Lett. {\bf 84}, 4757 (2000).
\bibitem{41}
J.~Blocki, J.~Randrup, W.~J.~Swiatecki,  C.~F.~Tsang,
Ann. Phys. (N.Y.) {\bf 105}, 427 (1977).
\bibitem{42}
B.~V.~Derjaguin, I.~I.~Abrikosova,  E.~M.~Lifshitz,
Q. Rev. Chem. Soc. {\bf 10}, 295 (1956).
\bibitem{43}
M.~Bordag, U.~Mohideen, V.~M.~Mostepanenko,
{ Phys. Rep.} {\bf 353}, 1 (2001).
\bibitem{51}
S.~G.~Rabinovich,
{\it Measurement Errors and Uncertainties.
Theory and Practice}
(Springer-Verlag, New York, 2000).
\bibitem{44}
G.~Bressi, G.\ Carugno, R.~Onofrio, G.~Ruoso,
Phys. Rev. Lett. {\bf 88}, 041804 (2002).
\bibitem{45}
C.~Yang, A.~Wax, R.~R.~Dasari, M.~S.~Feld,
Opt. Lett. {\bf 27}, 77 (2002).
\bibitem{49a}
S.~Brandt, {\it Statistical and Computational Methods in Data
Analysis} (North-Holland, Amsterdam, 1976).
\bibitem{46}
T.~Emig, R.~L.~Jaffe, M.~Kardar, A.~Scardicchio,
Phys. Rev. Lett. {\bf 96}, 080403 (2006).
\bibitem{47}
M.~Bordag,
Phys. Rev. D {\bf 73}, 125018 (2006).
\bibitem{48}
A.~Bulgac, P.~Magierski, A.~Wirzba,
Phys. Rev. D {\bf 73}, 025007 (2006).
\bibitem{49}
H.~Gies, K.~Klingm\"{u}ller,
Phys. Rev. Lett. {\bf 96}, 220401 (2006);
Phys. Rev. D {\bf 74}, 045002 (2006).
\bibitem{50}
D.~E.~Krause, R.~S.~Decca, D.~L\'opez, E.~Fischbach,
Phys. Rev. Lett. {\bf 98}, 050403 (2007).

\bibitem{52}
N.~W.~Ashcroft, N.~D.~Mermin,
{\it Solid State Physics}
(Saunders Colledge, Philadelphia, 1976).
\bibitem{53}
{\it American Institute of Physics Handbook}
(McGraw-Hill. New York, 1972).
\bibitem{54}
A.~Lambrecht, S.~Reynaud,
Eur. Phys. J. D {\bf 8}, 309  (2000).
\bibitem{55a}
E.~M.~Lifshitz, L.~P.~Pitaevskii,
{\it Statistical Physics}
(Pergamon Press, Oxford, 1980), Pt.~II.
\bibitem{55}
V.~B.~Bezerra,
G.~L.~Klimchitskaya, V.~M.~Mostepanenko,
 C.~Romero,
Phys. Rev. A {\bf 69}, 022119 (2004).
\bibitem {56}
C.~Genet, A.~Lambrecht,  S.~Reynaud,
Phys. Rev. A {\bf 62}, 012110 (2000).
\bibitem{57}
M.~Bordag, B.~Geyer, G.~L.~Klimchitskaya,
V.~M.~Mostepanenko,
{ Phys. Rev. Lett.}  {\bf 85}, 503  (2000);
   {\bf 87}, 259102  (2001).
\bibitem {58}
V.~B.~Bezerra, G.~L.~Klimchitskaya, C.~Romero,
   { Phys. Rev.} A
{\bf 65}, 012111 (2002).
\bibitem{59}
B.~Geyer, G.~L.~Klimchitskaya,
V.~M.~Mostepanenko,
Phys. Rev. A {\bf 67}, 062102 (2003).
\bibitem{60}
V.~B.~Bezerra,
R.~S.~Decca, E.~Fischbach, B.~Geyer,
G.~L.~Klimchitskaya, D.~E.~Krause,
D.~L\'opez,
V.~M.~Mostepanenko, C.~Romero,
{Phys. Rev. E} {\bf73}, 028101 (2006).
\bibitem{61}
J.~S.~H{\o}ye, I.~Brevik, J.~B.~Aarseth, K.~A.~Milton,
{J. Phys. A}: Math. Gen. {\bf 39}, 6031 (2006).
\bibitem {62}
L.~D.~Landau, E.~M.~Lifshitz, L.~P.~Pitaevskii,
{\it Electrodynamics of Continuous Media}
(Pergamon Press, Oxford,
1984).
\bibitem{63}
T.~Emig, A.~Hanke, R.~Golestanian, M.~Kardar,
Phys. Rev. Lett. {\bf 87}, 260402 (2001).
\bibitem {64}
C.~Genet, A.~Lambrecht, P.~Maia Neto, S.~Reynaud,
Europhys. Lett. {\bf 62}, 484 (2003).
\bibitem {65}
I.~Pirozhenko, A.~Lambrecht, V.~B.~Svetovoy,
New J. Phys. {\bf 8}, 238 (2006).
\bibitem{66}
D.~Iannuzzi, I.~Gelfand, M.~Lisanti, F.~Capasso,
in: {\it Quantum Field Theory under the Influence of
External Conditions}, ed. K.\ A.\ Milton
(Rinton Press, Princeton,2004).
\bibitem{67a}
S.~K.~Lamoreaux, W.~T.~Buttler,
Phys. Rev. E {\bf 71}, 036109 (2005).
\bibitem{67b}
M.~Brown-Hayes, D.~A.~R.~Dalvit, F.~D.~Mazzitelli,
W.~J.~Kim, R.~Onofrio,
Phys. Rev. A {\bf 72}, 052102 (2005).
\bibitem{67c}
P.~Antonini, G.~Bressi, G.~Carugno, G.~Galeazzi,
G.~Messineo, G.~Ruoso,
New J. Phys. {\bf 8}, 239 (2006).
\bibitem{67}
G.~L.~Klimchitskaya, F.~Chen, R.\ S.\ Decca,
E.\ Fischbach,  D.\ E.\ Krause, D.\ L\'opez,
U.~Mo\-hi\-deen, V.~M.~Mostepanenko,
J. Phys. A: Math. Gen. {\bf 39}, 6485 (2006).
\bibitem{68}
F.~Chen, U.~Mohideen, G.~L.~Klimchitskaya,
V.\ M.\ Mos\-te\-pa\-nen\-ko,
Phys. Rev. A  {\bf 74}, 022103 (2006).
\bibitem{69}
R.~S.~Decca, D.~L\'opez, H.~B.~Chan, E.~Fischbach,
D.~E.~Krause, C.~R.~Jamell,
{Phys. Rev. Lett.} {\bf 94}, 240401 (2005).
\end{thebibliography}
\end{document}